\documentclass[aps,prb,10pt,twocolumn,showpacs,superscriptaddress]{revtex4-1}

\newcommand{\sgn}{\mathop{\mathrm{sgn}}}
\usepackage{multirow}
\usepackage{color}
\usepackage{graphicx}
\usepackage{float}

\usepackage{dcolumn}

\begin{document}


\title{XPS core-level chemical shift by $ab~initio$ many-body theory}

\author{Iskander Mukatayev}
\affiliation{Universit\'e Grenoble Alpes, CEA, Leti, F-38000, Grenoble, France}

\author{Florient Moevus}
\affiliation{Universit\'e Grenoble Alpes, CEA, Leti, F-38000, Grenoble, France}

\author{Beno\^it Skl\'enard}
\affiliation{Universit\'e Grenoble Alpes, CEA, Leti, F-38000, Grenoble, France}
\affiliation{European Theoretical Spectroscopy Facility (ETSF)}

\author{Valerio Olevano}
\email{valerio.olevano@neel.cnrs.fr}
\affiliation{Universit\'e Grenoble Alpes, F-38000 Grenoble, France}
\affiliation{CNRS, Institut N\'eel, F-38042 Grenoble, France}
\affiliation{European Theoretical Spectroscopy Facility (ETSF)}

\author{Jing Li}
\email{jing.li@cea.fr}
\affiliation{Universit\'e Grenoble Alpes, CEA, Leti, F-38000, Grenoble, France}
\affiliation{European Theoretical Spectroscopy Facility (ETSF)}

\date{\today}

\begin{abstract}
X-ray photoemission spectroscopy (XPS) provides direct information on the atomic composition and stoichiometry by measuring core electron binding energies.
Moreover, according to the shift of the binding energy, so-called \textit{chemical shift}, the precise chemical type of bonds can be inferred, which brings additional information on the local structure.
In this work, we present a theoretical study of the chemical shift firstly by comparing different theories, from Hartree-Fock (HF) and density-functional theory (DFT) to many-body perturbation theory (MBPT) approaches like the $GW$ approximation and its static version (COHSEX). 
The accuracy of each theory is assessed by benchmarking against the experiment on the chemical shift of the carbon 1$s$ electron in a set of molecules.
More importantly, by decomposing the chemical shift into different contributions according to terms in the total Hamiltonian, the physical origin of the chemical shift is identified as classical electrostatics. 
\end{abstract}
\maketitle

\section{Introduction}

Since its discovery, X-ray photoemission spectroscopy (XPS) \cite{Siegbahn,Carlson} has been recognized as one of the most powerful experimental techniques, in both physics and chemistry, to characterize a broad range of systems, i.e. from 3D periodic crystals\cite{Bagus2013, Borgatti2018} and amorphous\cite{Saini2016, AarvaCaro19, AarvaCaro19a}, to 2D-materials\cite{Scardamaglia2017, Susi2018}, surfaces\cite{Giovannantonio2018} and superlattices\cite{sigesl}, up to molecules in gas or liquid phase.
Additionally, XPS provides direct information not only on the atomic structure, such as chemical composition and stoichiometry but also on the electronic structure, which could be accessed only indirectly by other techniques like optical absorption or energy-loss.

The atomic structure of a compound can be precisely established by XPS through the core-electron binding energies.\cite{Siegbahn,Carlson} 
The presence of a given atomic element is indicated in XPS spectra by the unique pattern of its core electron binding energies.
Meanwhile, the intensity of core electron XPS peaks, more precisely the relative ratio of peaks' intensities, associated with the various atomic components, delivers the actual stoichiometry of the compound, that is the chemical brute formula.
Finally, the shifts of core-level binding energies, so-called \textit{chemical shifts} provide valuable information about the chemical environment of the considered atomic element, that is, the type of chemical bonds.
Hence, XPS characterization offers a clear chemical picture of the compound with both brute chemical formula and structural information.

Although experimental core-electron binding energies are tabulated for the entire periodic table and experimental tables exist also for chemical shifts of core-electron binding energies for some important elements in several chemical environments, e.g.\ carbon, a complete experimental table of chemical shifts for any atom in any chemical environment is far to be available and achievable.\cite{AarvaCaro19,AarvaCaro19a}
A precise attribution of chemical shifts is complicated in compounds presenting a complex chemical structure.\cite{AarvaCaro19,AarvaCaro19a}
For example, a given atom could be bonded to many other elements, making it difficult to disentangle all the corresponding chemical shifts. 
The imperfect experimental situation, e.g.\ oxidation of the sample, presence of impurities, etc.\ raises the difficulty.
Therefore, an unambiguous attribution of the various chemical shifts is unattainable experimentally.
Fortunately, theory, especially first principles, could fill the missing part of the table for the chemical shift, which assists experimentalists in the analysis of XPS data.
However, attention has to be paid to the accuracy, as the necessary theoretical approximation introduces an unavoidable error bar.

In this work, we present a comprehensive study of the chemical shift which can be calculated using various \textit{ab initio} theories and approximations: from the simplest Hartree-Fock (HF) and density-functional theory (DFT) to the most advanced state-of-the-art many-body perturbation theory (MBPT) approaches to calculate electron binding/removal energies, e.g.\ quasiparticle energies, using the already well established $GW$ approximation\cite{Hedin65,StrinatiHanke80,StrinatiHanke82,HybertsenLouie85,GodbySham87}, whose validity is already well checked at least on valence electrons.
We also check the Coulomb-hole screened-exchange (COHSEX) approximation\cite{Hedin65}, a simplified static version of $GW$.
The validity and the accuracy of each theory are assessed by benchmarking against accurate measured chemical shifts for the 1$s$ core-electron binding energy of carbon in various molecules / chemical environments.
Additional energy decomposition analysis sheds light on the physical origin of the chemical shift, thus providing indications for simpler approaches, using less accurate and less cumbersome theories, to evaluate the chemical shift of elements at acceptable accuracy.

\begin{figure*}
\centering
\includegraphics{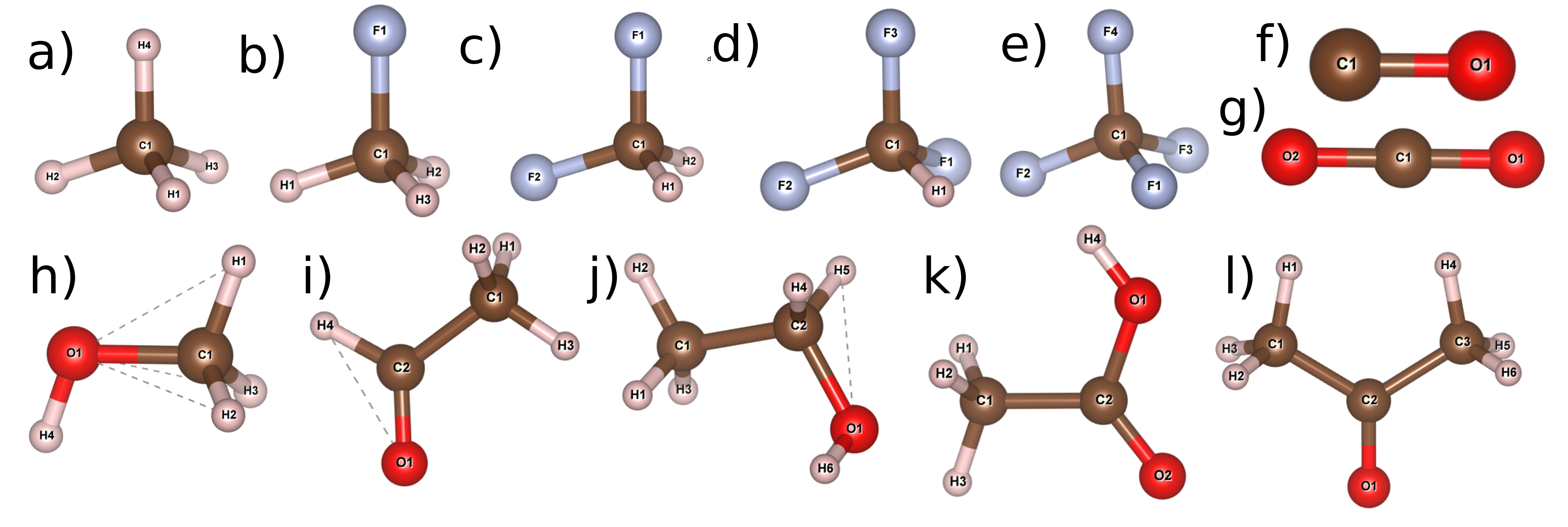}  
\caption{Structures of calculated molecules visualised using VESTA~\cite{Momma2011}: a) CH$_4$ methane, b) CH$_3$F methil-fluoride, c) CH$_2$F$_2$ dimethil-fluoride , d) CHF$_3$ trimethil-fluoride, e) CF$_4$ carbon-tetrafluoride, f) CO carbon-monoxide, g) CO$_2$ carbon-dioxide, h) CH$_3$OH methanol, i) CH$_3$COH acetal-aldehyde (ethanal), j) CH$_3$CH$_2$OH ethanol, k) CH$_3$COOH acetic acid, l) CH$_3$COCH$_3$ dimethil-ketone (propanone).}
\label{fig:molecules}
\end{figure*}

Previous works on core-level BEs have addressed the Delta self-consistent field ($\Delta$SCF) method within DFT \cite{JonesGunnarsson89,Martin}. This method is based on BEs calculations as the difference between total energies of neutral and ionized systems\cite{Bagus_1965}. These methods give deviations of chemical shifts approximately 0.2-0.3~eV\cite{PueyoBellafontIllas16} while the difference between absolute values of BEs and experiment is in several electronvolts. However, the calculation of periodic systems leads to the fact that the necessary ionization during the ($\Delta$SCF) generates a Coulomb divergence\cite{OzakiLee2017}, thus further approximations are required. Also the Delta coupled-cluster $\Delta$CC  gives highly accurate values of chemical shift and BEs\cite{ZhengCheng2019, Holme2011}, but the application of such a high-level method is limited by small systems, due to scaling factor with system size. 
Many-body calculations of core levels are much less present in the literature.
There are some exploratory studies on both solids\cite{IshiiOhno10,AokiOhno18} and molecules\cite{GolzeRinke18,VooraFurche19, VanSettenIllas18}, which have obtained partly promising results, but in some cases also large deviations from the experiment on absolute core-level BEs.
More recent works \cite{GolzeRinke20,MukatayevLi22} seems to have solved such problems.
However, the focus of the present work is specifically on the chemical shift.

\section{Systems}
We used the carbon 1$s$ core level as a reference for our theoretical study.
The C 1$s$ represents also the standard reference of almost all XPS experiments.\cite{Siegbahn,Carlson}
Indeed, its binding energy is used to fix the zero of the experimental energy scale, in particular, to circumvent in insulators the so-called \textit{charging} problem,\cite{Siegbahn,Carlson} which consists in a rigid shift of all energies simply caused by classical electrostatic charging of the sample.

To benchmark the theoretical prediction of the chemical shift, we used the set of molecules shown in Fig.~\ref{fig:molecules}.
This set contains various chemical environments (bonding) for carbon atoms, whose 1$s$ core-electron binding energies have been measured accurately by experiment. 
Chemical shifts of carbon 1$s$ are then obtained by taking methane (\textit{a} in Fig.~\ref{fig:molecules}) as a reference. 
Experimental molecular structures~\cite{Haynes2014-ip} were used in our calculations for all molecules of the set, i.e. without any geometrical relation. 
No other input has been taken from the experiment, the rest of the calculation is fully \textit{ab initio}.

\section{Methods}

The theoretical approach we used as reference is the mean-field Hartree-Fock method which, in isolated systems, already constitutes a good approximation or, in any case, a good starting point for corrections.
However, we preferred to refer to the more general hybrid PBEh($\alpha$) method\cite{AtallaScheffler13} which contains both HF at one side ($\alpha=1$) and the other very popular approach of density-functional theory (DFT) in the PBE\cite{LangrethMehl83,PerdewErnzerhof96} approximation at the other side ($\alpha=0$).
Since HF systematically overestimates electron binding energies (as well as HOMO-LUMO band gaps), whereas DFT-PBE systematically underestimates them, a judiciously chosen $\alpha$ adjustable parameter can provide BEs better in agreement with the experiment.
There is no rigorous derivation of the PBEh($\alpha$) approach.
However, this approach is much like in the spirit of the oldest Slater X$\alpha$ method that introduces correlations, which are always opposed in sign to exchange, by simply reducing the Fock exchange operator with a weighting coefficient $\alpha < 1$ in front of it.
This is the rationale of both Slater X$\alpha$ and PBEh($\alpha$).

We performed PBEh($\alpha$) all-electron calculations for a set of $\alpha$ between 0 and 1. We also considered the DFT local-density approximation (LDA), since this is a reference physical approximation of DFT.
Within all these approaches, we calculated both energies and wavefunctions for all the molecules of the considered set.
The C 1$s$ energy is the quantity we are focusing on, but we also considered the last occupied molecular orbital (HOMO) directly associated with the ionization potential (IP) of the molecule.
The reason is that on the IP a large literature of many other theoretical calculations is available for comparison, which is not the case for core states.

The HF, DFT-PBE, DFT-LDA or PBEh($\alpha$) electronic structure, both energies $\epsilon_n$ and wavefunctions $\psi_n(r)$, were used as starting points of our many-body perturbation theory (MBPT) calculations.
We first evaluate the Green function $G$:
\begin{equation}
  G(r, r', \omega) = \sum_n \frac{\psi_n(r)\psi_n(r')}{\omega - \epsilon_n -i\eta \sgn (\mu -\epsilon_n)}
  \label{G}
\end{equation}
where the sum runs over occupied and empty states, $\mu$ denotes the chemical potential, and $\eta$ is a positive infinitesimal.
From $G$ we evaluate the random-phase approximation (RPA) polarizability $\Pi = -iGG$, which at its turn enters into the screened Coulomb interaction $W = v + v\Pi W$, with $v = 1/|r-r'|$ the bare Coulomb interaction.
We have now all the ingredients to calculate the $GW$ self-energy,
\begin{equation}
  \Sigma(r,r',\omega) = \frac{i}{2\pi}\int\,d\omega'e^{i\omega'\eta}G(r,r',\omega+\omega')W(r,r',\omega')
  \label{sigma}
\end{equation}
where the integral is carried on by contour deformation.
Finally, quasiparticle energies $\epsilon^\mathrm{QP}_n$ can be calculated to first order perturbation theory by the equation
\begin{equation}
  \epsilon^\mathrm{QP}_n = \epsilon_n + \Re \langle\psi_n|\Sigma(\epsilon^\mathrm{QP}_n) - V_{xc} |\psi_n\rangle
  , \label{epsilon}
\end{equation}
where $\epsilon_n$ and $\psi_n$ are the energy and wavefunction zero-order starting approximation (HF, PBE, PBEh) and $V_{xc}$ is the corresponding exchange-correlation potential.
The binding energy of an electron is its quasiparticle energy with a reversed sign.
The calculation procedure can stop here, and this is called non self-consistent $G_0W_0$; or the new QP electronic structure can be re-injected into the Green function Eq.~(\ref{G}) and the procedure reiterated until self-consistency is achieved.
The self-consistency we have considered here is based on a re-injection of only energies, not wavefunctions (ev$GW$).
Alternatively, we have considered a full self-consistency, both energies and wavefunctions, within the so-called Coulomb-hole screened exchange (COHSEX) approximation\cite{Hedin65}, which is a static version of the $GW$ approximation.

Calculations were carried out using the aug-cc-pVQZ augmented correlation-consistent Gaussian basis set, and the auxiliary basis set def2-Universal-JKFIT \cite{Weigend_2008} for the Coulomb-fitting resolution of the identity (RI-V)\cite{Duchemin_2017}.
We used the \textsc{NWChem} code\cite{Valiev_2010} for HF and DFT calculations, and the \textsc{Fiesta} code\cite{BlaseOlevano11,Jac15a,Li16} for $GW$ and COHSEX.

\section{Results and Discussion}

\begin{figure}[t]
\centering
\includegraphics[width=\columnwidth]{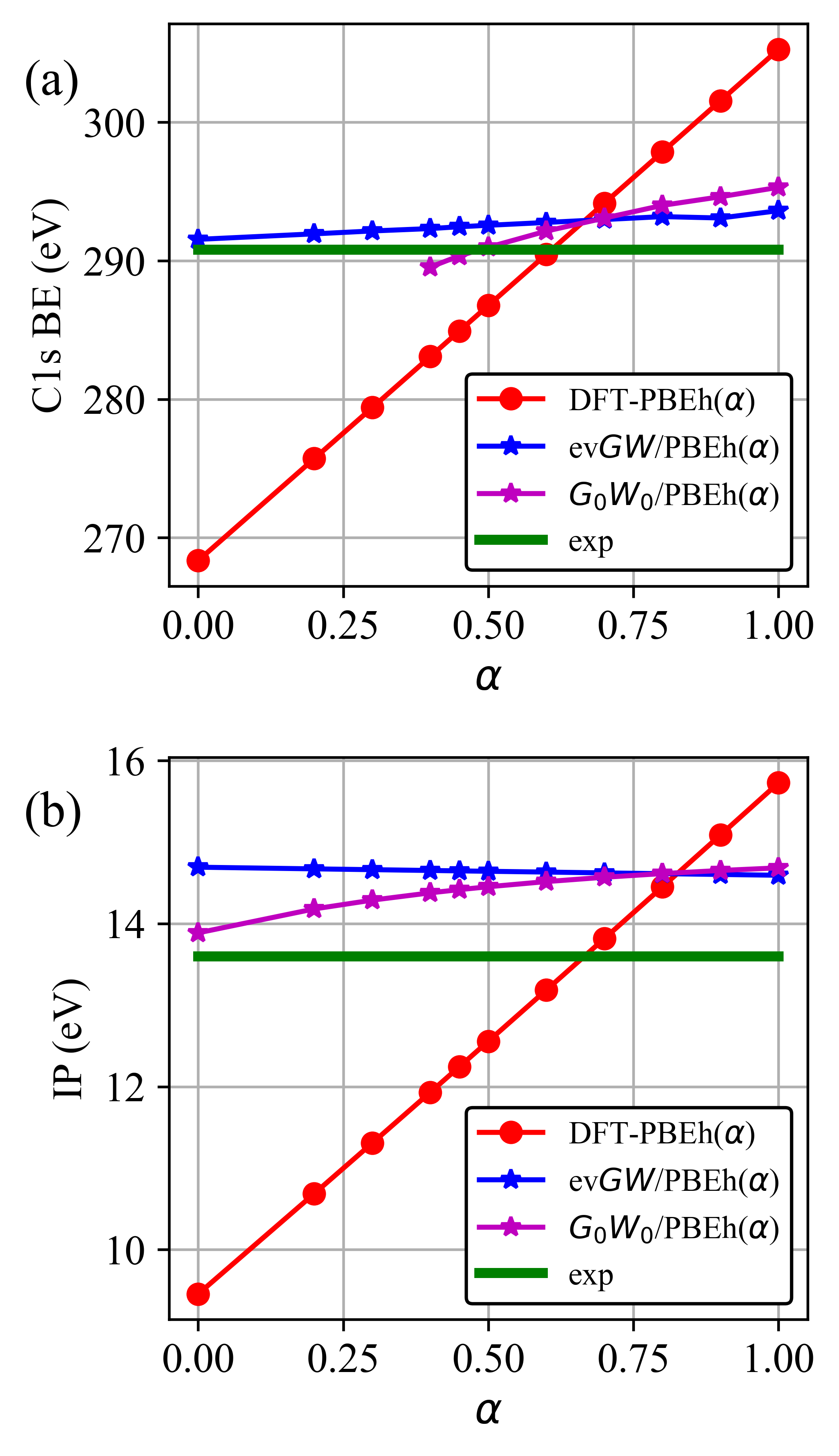}  
\caption{Methane CH$_4$ molecule electron BE for the C 1$s$ level (a) and IP (b).
Red curve and circles: PBEh($\alpha$); violet curve and stars: $G_0W_0$/PBEh($\alpha$);
blue curve and stars: ev$GW$/PBEh($\alpha$); green horizontal lines: experimental values.}
\label{fig:imag1}
\end{figure}

\subsection{C 1$s$ absolute energy vs ionization potential:\\
HF, PBE, hybrids and $GW$ correlation effects}

Before tackling the chemical shift, which is our main focus in this work, we have first studied the C 1$s$ core level absolute energy in CH$_4$ methane, the reference molecule for chemical shifts.
In this preliminary study, we compared the performances of the various approximations on the C 1$s$ core level absolute binding energy (BE) compared to the ionization potential (IP), a quantity over which there is much more understanding and literature.

\begin{figure*}
\centering
\includegraphics{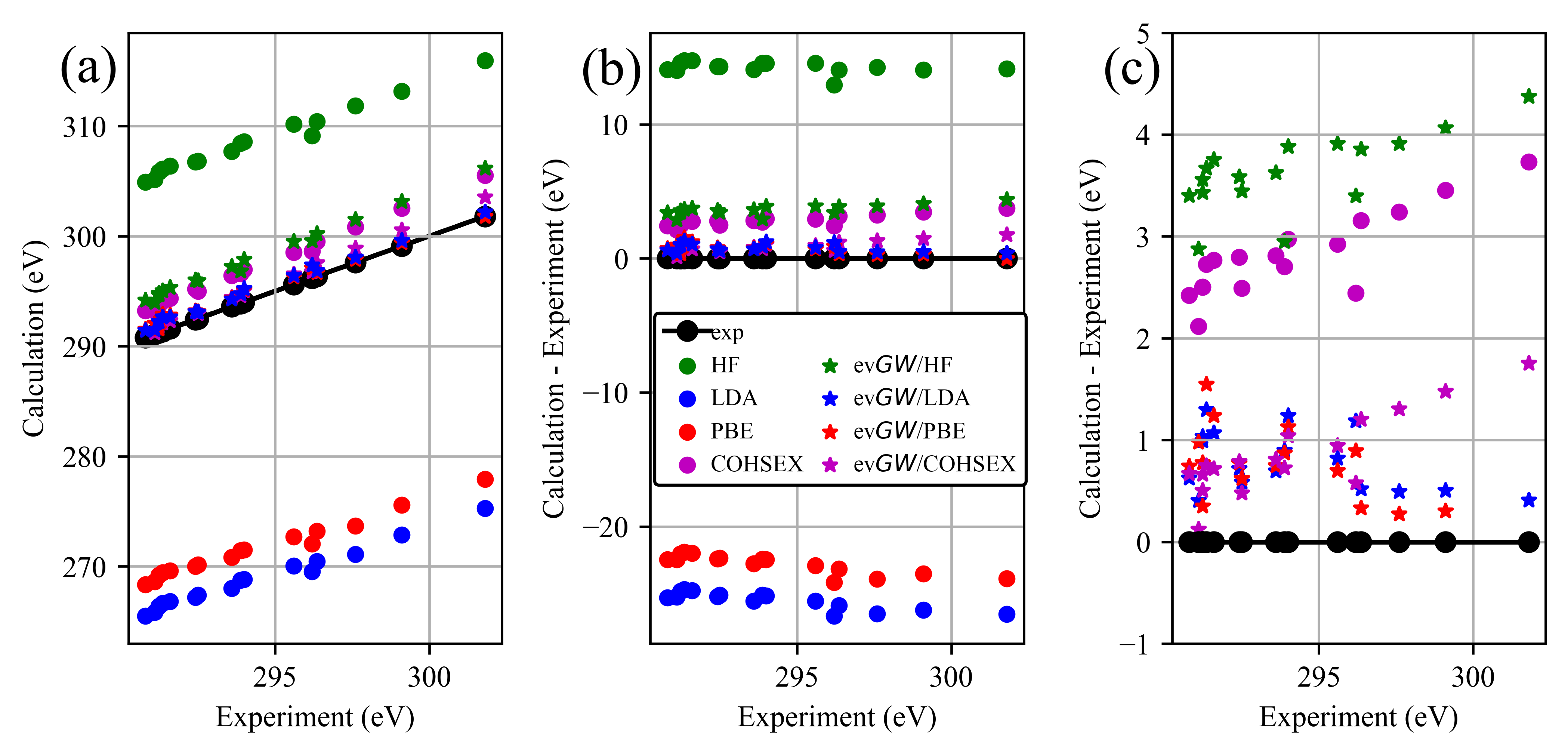}  
\caption{C 1$s$ electron BE for the set of molecules in Fig.~\ref{fig:imag1} computed with different methods versus the experiment: (a) absolute values; (b) error with respect to the experiment; (c) zoom around 0 of (b).}
\label{fig:imag2}
\end{figure*}

This comparison is presented in Fig.~\ref{fig:imag1}: the top panel (a) for the methane C 1$s$ core level absolute BE, and the lower panel (b) for the methane IP, equal to the binding energy of the highest occupied molecular orbital (HOMO).
Green horizontal lines are placed exactly at the experimental values.
The red curves present the results obtained for the hybrid functional PBEh using various values of the adjustable mixing parameter $\alpha$, from 0 to 1.
The same red curve presents also the results for DFT-PBE and HF-like functional: the red dot at $\alpha=0$ corresponds to the result obtained in DFT PBE, whereas that one at $\alpha=1$ corresponds to the result obtained in HF-like functional (exact-exchange with  PBE correlation).
It can be seen that HF and DFT PBE have the same behaviour on the core level BE as well as on the IP.
HF underestimates both the core level BE and the IP, whereas DFT PBE overestimates both.
The absolute error is larger in the core level BE, but this is not the case if we consider the relative error.
Since the hybrid PBEh constitutes a linear interpolation between PBE ($\alpha=0$) and HF ($\alpha=1$) which respectively underestimates and overestimates the experiment, there exists of course a value of $\alpha$ which exactly reproduces the experiment.
We can read it in the figure as the intercept of the red and the green lines.
The problem is that this value of $\alpha$ is different if we consider the IP or the C 1$s$ BE : $\alpha = 0.6$ in the latter and $\alpha=0.7$ in the former case.
This shows that the hybrid PBEh is just only a phenomenological approximation, driven by common sense, but without a more in-deep physical justification.
It can be taken as an adjustable parameter approach which, in any case, is superior to a simplified model because the Hamiltonian is still microscopic.

In the same Fig.~\ref{fig:imag1}, we report the results obtained by single iteration $G_0W_0$ (violet curve) and the eigenvalue self-consistent ev$GW$ (blue curve) approximation using as starting point PBEh at the various $\alpha$, including of course also the $\alpha=1$ HF and the $\alpha=0$ DFT PBE cases.
The first thing to be remarked is the fact that the dependence of the $GW$ result on the $\alpha$ adjustable parameter is much reduced.
This is already the case for a single iteration of $GW$ correction, that is $G_0W_0$.
The $\alpha$ dependence becomes almost absent in eigenvalue self-consistent ev$GW$.
The ev$GW$ IP value is practically independent of $\alpha$.
On the C 1$s$ BE there is still a residual dependence on $\alpha$ even in ev$GW$, which indicates that wavefunctions matter more in core levels than in the IP.
This points to the need for core levels of many-body corrections also to wavefunctions, beyond only energies.

When we analyze the IP, we see that although the $G_0W_0$ result presents a dependence on the starting $\alpha$, an $\alpha$ able to reproduce the experimental result could not be found, unless to use negative $\alpha$s which is even less justifiable from a physical point of view.
On the other hand, the ev$GW$ IP is almost perfectly constant, which would be a good point, weren't for the fact that the obtained constant is $\sim$1 eV higher than the experimental value.
Surprisingly, the experimental IP of methane could not be reproduced accurately by $GW$.
The IP of methane we found is consistent with previous results reported in literature \cite{Maggio2017}.

As to the C 1$s$ core level BE, the analysis of the $G_0W_0$ correction on top of PBEh shows that 1) the result depends on $\alpha$; 2) that at $\alpha=0.45$ the experimental result of 290.8~eV \cite{Bakke1980} is recovered; 3) that for $\alpha<0.4$ Eq.~\ref{epsilon} does not present any more a clear quasiparticle energy solution.
This is a drawback that manifests whenever the plasmon energy, or the main pole of the screening function $\varepsilon^{-1}(\omega)$, is comparable to the $G_0W_0$ energy correction.
This problem on core levels was already reported by Golze et al~\cite{GolzeRinke20}.
However, as Fig.~\ref{fig:imag1}(b) shows clearly, it can be resolved by changing the start point.
For example, choosing HF ($\alpha=1$), or hybrids around ($\alpha>0.4$) as starting point, instead of PBE in these cases for core levels.
We finally remark that, like for the IP, the ev$GW$ overestimates the C 1$s$ BE by more than 1~eV.
The ev$GW$ result is more sensitive to the starting $\alpha$ in this case than in the case of the IP.
However, there is no $\alpha$ able to reproduce the experimental BE.
This points again to the limited physical validity of the PBEh approach because of the absence of a universal $\alpha$ which could represent a compromise between the various observables/quantities to be reproduced at the same time. 
For this reason, in our study of the C 1$s$ chemical shift, instead of sticking on PBEh with a fixed $\alpha$, we prefer to continue to bring standard functionals (LDA, PBE, or HF) as starting point for $GW$.

\subsection{C 1$s$ chemical shift}
The C 1$s$ chemical shift is studied for the set of molecules shown in Fig.~\ref{fig:molecules} and presented in Fig.~\ref{fig:imag2}.
In the latter, we report the results obtained for DFT-LDA, DFT-PBE, HF, COHSEX, as well as ev$GW$ on top of all previous approaches, with respect to the experiment~\cite{Bakke1980} on the abscissa.
Panel (a) presents the absolute BE, whereas in (b) and in its zoom (c), the theory minus experiment error is presented.
We again notice the large underestimation by more than 24~eV of both PBE and LDA; the overestimation by about 12~eV of HF; and the improvement obtained by applying ev$GW$ on top of them.
ev$GW$ always converges toward the experiment, no matter if the starting result is overestimating or underestimating it.
We then remark the already good result obtained in COHSEX with an overestimation of only 3~eV, which is further improved when applying ev$GW$ on top of it.

\begin{figure}
\centering
\includegraphics{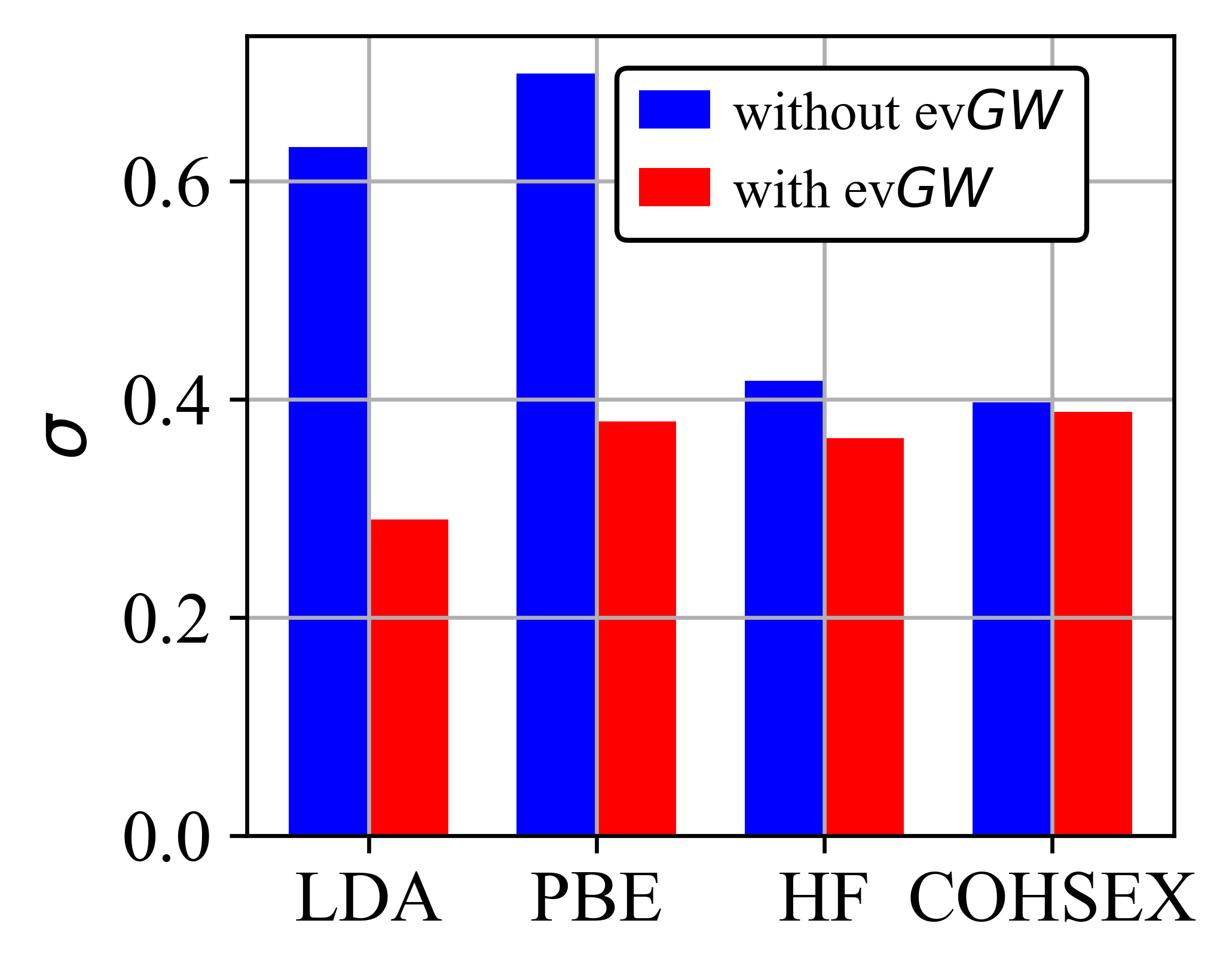}  
\caption{Variances on the C 1$s$ chemical shift computed for different methods with respect to the experiment.}
\label{fig:imag3}
\end{figure}

The chemical shift is the difference of a given C 1$s$ BE in a molecule with respect to the C 1$s$ BE of the reference molecule, that is methane. 
So, the chemical shift is relative energy. 
What matters for a given approach to have a correct reproduction of the chemical shift, is not the absolute error from the experimental BE for a given molecule, but rather that this error is constant all along the different molecules and chemical environments.
An approach that performs correctly on the chemical shift should present in Fig.~\ref{fig:imag2}(a) a curve that is as much as possible parallel to the experimental curve, or as much as possible horizontal in 
Fig.~\ref{fig:imag2}(b) or (c).
But looking at these figures, we see that this is more or less the case for all the considered methods: they are all scattered along almost horizontal lines in Fig.~\ref{fig:imag2}b, even the LDA and PBE approaches which, we have already seen, present the largest absolute underestimations and errors.
To have a more clear picture of the performances of the various approaches, we computed for them the variance specifically for the chemical shift observable, which is reported in Fig~\ref{fig:imag3}.
Thus, while the chemical shift spans almost 15~eV, the variance in both LDA and PBE is only about $0.65$ eV.
This implies that most of the chemical shift is already captured at the DFT level.
From the same figure, we see that HF and COHSEX, with a variance of 0.4~eV, perform better than DFT.
In all cases, the application of ev$GW$ improves the result and achieves a variance of only 0.3~eV for the case of ev$GW$ on top of LDA.

\begin{figure}
\centering
\includegraphics{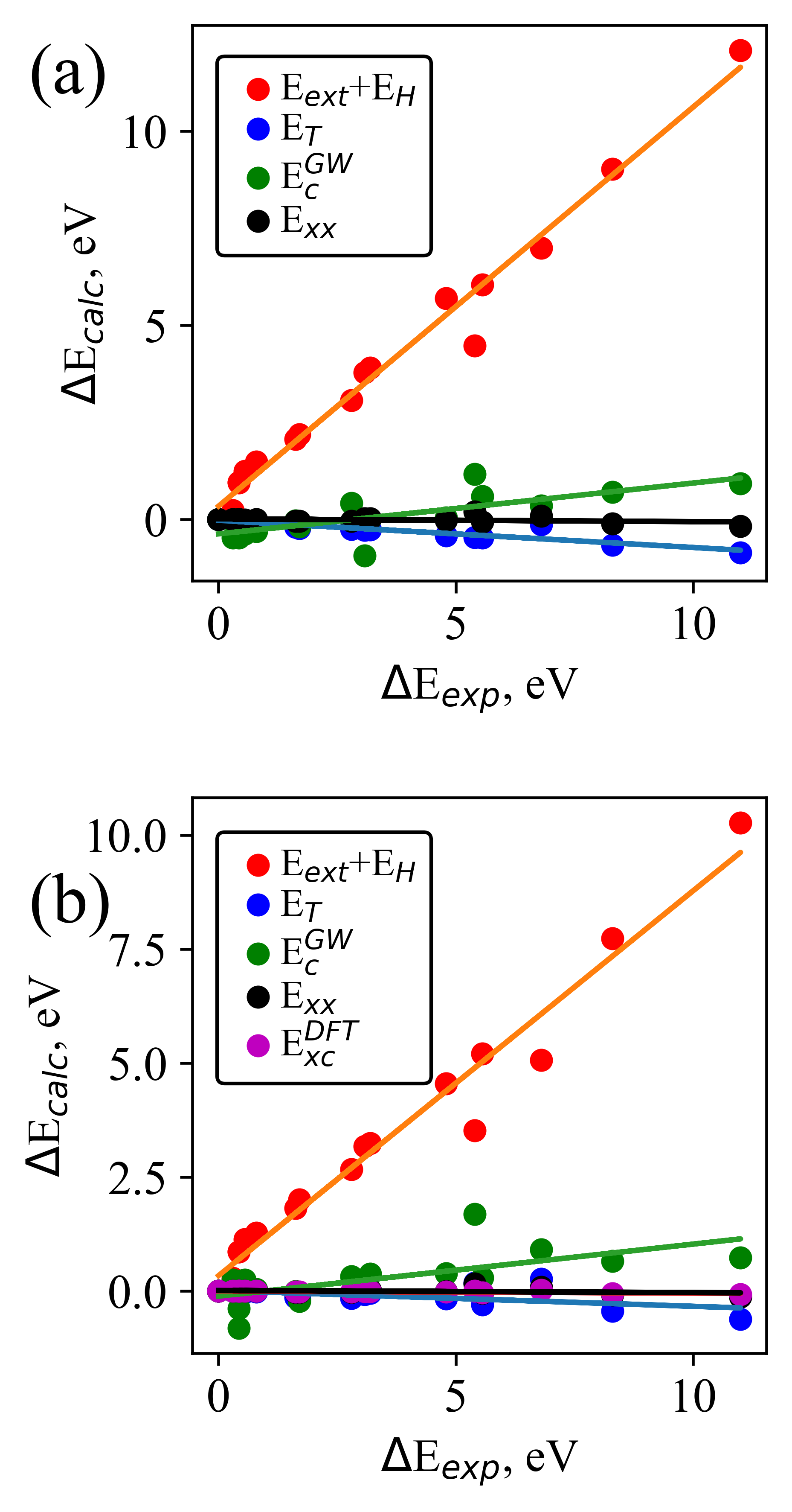}  
\caption{Energy decomposition of C 1$s$ chemical shift in all its contributions for HF and $GW$ on top of HF (a); and PBE and $GW$ on top of PBE (b).}
\label{fig:imag4}
\end{figure}

\subsection{Origin of chemical shift}

The surprising result that the chemical shift does not require much theoretical complexity and it is already well captured at the level of DFT, awaits an explanation.
To reveal the origin of the chemical shift, we studied the decomposition of the C 1$s$ level in all its energy contributions.
The quasiparticle energy of an electron state $n$
\[
\epsilon_n = \langle \psi_n | \hat{H} | \psi_n \rangle
\]
can be split into all the terms composing the quasiparticle Hamiltonian
\[
 \hat{H} = \hat{T} + \hat{V}_\mathrm{ext} + \hat{V}_\mathrm{H} + \hat{\Sigma}_\mathrm{xc},`
\]
where $\hat{T}$ is the kinetic operator, $\hat{V}_\mathrm{ext}$ the external potential due to the ions, and $\hat{V}_\mathrm{H}$ the classical Hartree potential from electron repulsion.
$\Sigma_\mathrm{xc}$ is the exchange-correlation term that depends on the chosen approximation.
In DFT, it is the LDA or PBE (or else) exchange-correlation potential  $v_\mathrm{xc}$; in HF the Fock exchange operator $\Sigma_\mathrm{x}$; in $GW$ $\Sigma_\mathrm{xc}$ can be further decomposed into the exchange and the $GW$ correlation operator, $\Sigma_\mathrm{xc} = \Sigma_\mathrm{x} + \Sigma^{GW}_\mathrm{c}$.
All these operators decompose the quasiparticle energy into the corresponding terms $\epsilon_\mathrm{T}$, $\epsilon_\mathrm{ext}$, $\epsilon_\mathrm{H}$, $\epsilon^\mathrm{DFT}_\mathrm{xc}$, $\epsilon_\mathrm{x}$, $\epsilon^{GW}_\mathrm{c}$.
The C 1$s$ chemical shift of a molecule from the CH$_4$ methane reference,
\[
 E^\mathrm{mol} = \epsilon^\mathrm{mol} - \epsilon^\mathrm{CH_4} 
\]
can also be decomposed in all its contributions, $E_\mathrm{T}$, $E_\mathrm{ext}$, $E_\mathrm{H}$, $E^\mathrm{DFT}_\mathrm{xc}$, $E_\mathrm{x}$, $E^{GW}_\mathrm{c}$, and these are reported in Fig.~\ref{fig:imag4}.
The $E_\mathrm{ext}$ and the $E_\mathrm{H}$ are in absolute the largest contributions, but since they are opposite in sign and the classical electrostatic attraction and repulsion that they represent largely balance each other, we decided to group them in only one \textit{classical} term.
Although merged, it is evident from Fig.~\ref{fig:imag4} that the classical electrostatic term $E_\mathrm{ext} + E_\mathrm{H}$ (red dots) continues to be the dominant contribution to the chemical shift, and this both in the straightforward HF and PBE cases, with no changes even if we correct them by $GW$ correlations.
We can already conclude that \textit{the chemical shift}, felt by an atomic core level due to the various chemical environments, \textit{is inherently a purely electrostatic classical effect not requiring a precise complex quantum description including exchange and correlation.}

\begin{table}[t]
\centering
\begin{tabular}{l c c c c c}
0th approx & $E_\mathrm{T}$ & $E_\mathrm{ext}+E_\mathrm{H}$& $E_\mathrm{x}$ & $E_\mathrm{xc}^\mathrm{DFT}$ & $E_\mathrm{c}^\mathrm{GW}$ \\
\hline
HF & -0.069 & 1.026 & -0.007 & & 0.135 \\
PBE & -0.036 & 0.843 & & -0.005 & 0.117 \\
\end{tabular}
\caption{C 1$s$ chemical shift slope: decomposition into its various contributions.}
\label{tab:table1}
\end{table}

To better understand, for each contribution we performed linear fits with respect to the experimental chemical shift, and these are represented in Fig.~\ref{fig:imag4} as straight lines, while their slopes are reported in Table~\ref{tab:table1}.
From the Table, we can see that the classical electrostatic contribution is by far the largest contribution.
Surprisingly, the $GW$ correlation term is the second-largest contribution, although already one order of magnitude smaller than the classical electrostatic contribution.
The remaining terms are negligible: the kinetic contribution is two orders of magnitude smaller, while both the HF exchange and the DFT exchange-correlation contributions are even three orders of magnitude smaller.

Interestingly, looking again at Fig.~\ref{fig:imag4}, we remark that at about 5.5~eV for the experimental chemical shift, a red dot falls outside the straight line fit: in both the HF and the PBE cases.
This dot corresponds to the CO molecule.
This is the only case where the classical electrostatic contribution is not enough to fully describe the chemical shift, and a missing important contribution is required which is surprisingly not the kinetic, nor the exchange, but the correlation contribution (green dot in Fig.~\ref{fig:imag4}).
In Fig.~\ref{fig:imag4}b we notice another red point falling below the linear fit at about 7~eV for the experimental chemical shift. However, it is not the case in Fig.~\ref{fig:imag4}a: this point corresponds to the CO$_2$ molecule.
In this case, the missing contribution seems to be again the $GW$ correlation term.
A possible explanation of this anomaly could be the fact that the two evidenced molecules present resonance chemical structures.
For example, the CO molecule is described to exist as the superposition of three resonances with single, double, and triple bonds of the carbon atom with oxygen.
This quantum superposition cannot be described classically and requires the inclusion of correlation effects.
We mention that a similar explanation was already provided\cite{Siegbahn113} for the case of another anomalous behaviour found for molecules such as CO and CO$_2$.

Therefore we can further conclude that, although the classical electrostatic term is the dominant contribution, correlations, as brought by the $GW$ approximation, improve the description of the chemical shift in particular cases.
Kinetic and exchange contributions can always be safely neglected in chemical shifts.
Same conclusion for the exchange-correlation contribution brought by the simplest DFT approximations.

\section{Conclusions}
We have employed several theoretical \textit{ab initio} approaches to calculate the C 1$s$ core level energy and its chemical shift compared with experimental data on a benchmark set of molecules.
Regarding absolute core level binding energies, the $GW$ approximation, especially in its eigenvalue-only partial self-consistency flavor, provides important improvements on both the HF overestimation and the DFT LDA and PBE underestimation. 
When using the single iteration non self-consistent $G_0W_0$, it is preferable to start with HF as the first trial, or it might be impossible to find a solution to the QP equation.
COHSEX is already a good approximation for core level binding energies, but $GW$ on top of COHSEX is the same quality as $GW$ on top of any other zero-order approximation.
Although it is likely possible to find in straightforward PBEh an $\alpha$ able to provide a result in perfect agreement with the experiment (but the $\alpha$ found is different for different observables), this is not the case when applying $GW$ on top of PBEh: the final result depends weakly or does not depend at all, on the $\alpha$ of the zero-order PBEh.

On the other hand, the chemical shift is dominated by the classical electrostatic (external plus Hartree) contribution, implying that low-level theoretical approaches can already describe the chemical shift with acceptable accuracy.
Correlation effects, as brought by the $GW$ approximation, are needed only in particular cases, as those illustrated in the present work, associated with the presence of quantum resonances like in the CO and CO$_2$ molecules.

\section{Acknowledgements}
Part of the calculations were using the allocation of computational resources from GENCI–IDRIS (Grant 2021-A0110912036).

\bibliography{ref,nobles}
\end{document}